\documentclass[11pt,a4paper]{JHEP3}
\usepackage{amsfonts,ulem,bbold}   
\usepackage{amssymb,amsmath}

\newcommand{\be}{\begin{equation}}  
\newcommand{\ee}{\end{equation}}  
\newcommand{\bea}{\begin{eqnarray}}  
\newcommand{\eea}{\end{eqnarray}}

\newcommand{\tf}{{}^{{}^{{}_{3}}}\!\!f}

\DeclareMathOperator{\tr}{tr}

\title{Bimetric Gravity from Ghost-free Massive Gravity}  

\author{S. F. Hassan\\ Department of Physics \& The Oskar  
Klein Centre,\\ Stockholm University, AlbaNova University Centre,  
SE-106 91 Stockholm, Sweden \\ E-mail: \email{fawad@fysik.su.se}
}  
  
\author{Rachel A. Rosen\\ Physics Department and Institute for
  Strings, Cosmology, and Astroparticle Physics,\\ 
Columbia University, New York, NY 10027, USA \\ E-mail:
\email{rar2172@columbia.edu}}  
  
\abstract{Generically, non-linear bimetric theories of gravity suffer from the same Boulware-Deser ghost instability as non-linear theories of massive gravity.  However, recently proposed theories of massive gravity have been shown to be ghost-free.  These theories are formulated with respect to a flat, non-dynamical reference metric.  In this work we show that it is possible to give dynamics to the reference metric in such a way that the consistency of the theory is maintained.  The result is a non-linear bimetric theory of a massless spin-2 field interacting with a massive spin-2 field that is free of the Boulware-Deser ghost.  To our knowledge, this is the first construction of such a ghost-free bimetric theory.}


\preprint{}

\begin{document}
  
\section{Introduction and summary} 
Generically, non-linear bimetric theories of gravity \cite{ISS} suffer from the same Boulware-Deser ghost instability \cite{BD} as non-linear theories of massive gravity. In this paper we construct non-linear bimetric actions that are free of the Boulware-Deser ghost.  This is possible due to the recent progress made in constructing ghost free theories of massive gravity.

Until recently, consistent non-linear extensions of Fierz-Pauli
massive gravity \cite{FP1,FP2} remained elusive.  In \cite{dRG2,dRGT}
a two-parameter family of massive gravity actions was proposed and
shown to be free of the Boulware-Deser \cite{BD} ghost in the
so-called decoupling limit and more generally at low order in
perturbation theory.  The absence of the Boulware-Deser ghost was
demonstrated at the full non-linear level in \cite{HR3}, based on the
reformulation of the theory given in \cite{HR1}. (For complementary
analyses see \cite{dRGT3}, \cite{dRGT4}. For a review of massive
gravity and the associated literature, see \cite{Hinterbichler}.)

To construct non-linear generalizations of the Fierz-Pauli mass term,
an additional ``reference" metric $f_{\mu \nu}$ is invariably
required. In the works mentioned above, this reference metric is taken
to be flat and non-dynamical.  There are many reasons one would like to
go beyond a flat, non-dynamical $f_{\mu \nu}$.  From a theoretical standpoint, promoting
$f_{\mu \nu}$ to a full dynamical metric is desirable as leads to a background-independent theory which is
invariant under general coordinate transformations, without the
introduction of St\"uckelberg fields.

Such theories were introduced in \cite{ISS} in order to describe the
interaction of gravity with a massive spin-2 meson.  More recently,
there has been renewed interest in bimetric theories of gravity due to
their accelerating cosmological solutions (see, e.g.,
\cite{DKP}). However, consistent theories of interacting spin-2 fields
have proved difficult to obtain \cite{AD1,AD2} for the following
reason. 

Generically, bimetric theories give rise to one massive
and one massless spin-2 field \cite{ISS,BD,SS,DK}.  In fact, it was
shown in \cite{BDGH} that there are no consistent theories of
interacting massless spin-2 fields.  Thus theories with multiple
interacting spin-2 fields will necessarily include a massive spin-2
field.  As a result, bimetric theories are plagued with the same ghost
problem that generically afflicts theories of massive gravity
\cite{BD,SS,DK}. Thus, ghost-free massive gravity is a promising
starting point from which to develop a consistent theory of
interacting spin-2 fields.

In order to consistently promote the reference metric $f_{\mu\nu}$ of
a massive gravity theory to a dynamical variable, it is first
necessary to show that the massive gravity theory remains ghost free for a
generic non-flat, non-dynamical $f_{\mu \nu}$.  This would guarantee that
fluctuations of $f_{\mu \nu}$ will not disrupt the consistency of the
$g_{\mu \nu}$ sector of the theory.  Such a proof was carried out in
\cite{HR4}.  Next, one must introduce a kinetic term for $f_{\mu \nu}$
in such a way that the $f_{\mu \nu}$ sector of the theory is also
ghost-free.  This is the aim of the present work.

We start by reviewing the recently proposed massive gravity actions
formulated with respect to a non-dynamical reference metric.  We then
show that the ghost-free mass term for the original metric $g_{\mu
  \nu}$ can be reformulated as a ghost-free mass term for $f_{\mu
  \nu}$.  This implies that the correct, consistent kinetic term for
$f_{\mu \nu}$ is in fact the Einstein-Hilbert term.  We next consider
the minimal bimetric model. When expanded around a fixed background,
we show that one linear combination of fluctuations of the metrics
describes a massive spin-2 field while the other combination describes
a massless spin-2 field.  We go on to discuss the Hamiltonian and
momentum constraints of the general bimetric theory. First we give general
arguments that the full non-linear theory also describes one massive
spin-2 field interacting with a massless spin-2 field.  We then
consider the minimal model in the ADM formulation at the full
non-linear level. We solve the constraints explicitly and demonstrate
that our general counting arguments go through. We then extend these results to the most general bimetric model.  We end with a brief
discussion of the coupling of the bimetric theory to matter.

\section{Non-linear massive gravity} 

The most general ghost-free non-linear massive gravity theories are
given by a two-parameter family of actions. For a flat reference
metric, these were proposed in \cite{dRG2,dRGT} and shown to be ghost
free at the complete non-linear level in \cite{HR3}. We are interested in the generalization of these
models to arbitrary $f_{\mu\nu}$ which was also shown to be ghost
free in \cite{HR4} at the complete non-linear level. For our 
purposes it is convenient to work with the reformulation of the
massive action given in \cite{HR1} for a general $f_{\mu\nu}$,  
\be
\label{act2}   
S=M_p^2\int d^4x\sqrt{-g}\,\bigg[R +2m^2 \sum_{n=0}^{4} \beta_n\,
  e_n(\sqrt{g^{-1} f})\bigg] , 
\ee 
where the square root of the matrix is defined such that
$\sqrt{g^{-1}f} \sqrt{g^{-1}f} = g^{\mu \lambda}f_{\lambda \nu}$.  The
$e_k(\sqrt{g^{-1} f})$ are elementary symmetric polynomials of the
eigenvalues $\lambda_n$ of the matrix $\sqrt{g^{-1} f}$.  For a $4
\times 4$ matrix they can be written as,
\bea
\label{e1}
e_0(\sqrt{g^{-1} f})&=&1 , \\[.1cm] 
e_1(\sqrt{g^{-1} f})&=&\lambda_1+\lambda_2+\lambda_3+\lambda_4 \, ,
\nonumber \\[.1cm]  
e_2(\sqrt{g^{-1} f})&=&\lambda_1 \lambda_2 + \lambda_1 \lambda_3+
\lambda_1 \lambda_4  + \lambda_2 \lambda_3 + \lambda_2 \lambda_4 +
\lambda_3 \lambda_4 , \nonumber \\[.1cm]  
e_3(\sqrt{g^{-1} f})&=& \lambda_1 \lambda_2 \lambda_3+\lambda_1
\lambda_2 \lambda_4+\lambda_1 \lambda_3 \lambda_4+ \lambda_2 \lambda_3
\lambda_4   \, , \nonumber \\[.1cm]  
e_4(\sqrt{g^{-1} f})&=&\lambda_1\lambda_2 \lambda_3 \lambda_4 = \det
\sqrt{g^{-1}f} \, . \nonumber  
\eea
The highest order term in the potential in (\ref{act2}), 
\be
\label{e4}
\sqrt{-\det g}\, \beta_4\, e_4(\sqrt{g^{-1} f}) =  \beta_4\,\sqrt{-\det f}\,, 
\ee
is independent of $g_{\mu \nu}$ and so does not contribute to the
$g_{\mu \nu}$ equations of motion.  Thus when the reference metric
$f_{\mu \nu}$ is taken to be non-dynamical, this term can be
neglected.  The four remaining $\beta_n$ describe four combinations of
the mass of the graviton, the cosmological constant and two free
parameters.

In what follows, we wish to consider the case that $f_{\mu \nu}$ is
dynamical and thus the potential term in (\ref{act2}) acts as a
potential for $f_{\mu \nu}$ as well as $g_{\mu \nu}$.  Thus the term
(\ref{e4}) should be retained in the action.  The resulting theory has
five free parameters, including the graviton mass and the cosmological
constants for both $g_{\mu \nu}$ and $f_{\mu \nu}$.

In ref. \cite{HR4}, the action (\ref{act2}) was analyzed in the ADM
formulation for a general non-flat but non-dynamical $f_{\mu \nu}$.
In the ADM formulation, there are six potentially dynamical modes described
by $g_{ij}$ and their conjugate momenta $\pi^{ij}$.  It was shown in \cite{HR4} and \cite{HR6} that, for these theories (\ref{act2}),
there is a Hamiltonian constraint along with an associated secondary
constraint.  These constraints remove one
propagating mode from the theory, leaving only the five modes
consistent with a massive spin-2 particle.  Thus, for a non-dynamical
$f_{\mu\nu}$, the $g_{\mu \nu}$ sector of theory is free of the
pathological sixth mode known as the Boulware-Deser ghost.  The
existence of the constraint is due to the specific form of the mass
term in (\ref{act2}). 

\section{Bimetric gravity}

Let us now consider the possibility of adding a kinetic term for $f_{\mu\nu}$.  For a dynamical $f_{\mu\nu}$, the mass term in (\ref{act2}) acts as a potential for $f_{\mu \nu}$ as well as for $g_{\mu \nu}$.  We want to add a kinetic term for $f_{\mu\nu}$ that is consistent with this potential, to get a ghost-free theory.  The given mass term does not appear to have the same form for $f_{\mu \nu}$ as it does for $g_{\mu \nu}$.  This is not automatically calamitous, however, as we are not constrained to write only the Einstein-Hilbert kinetic term for $f_{\mu \nu}$. In general, kinetic terms for $f_{\mu \nu}$ can be constructed using both metrics $g_{\mu \nu}$ and $f_{\mu \nu}$.  It could be that there exist multiple kinetic terms that give rise to a ghost-free theory in the $f_{\mu \nu}$ sector. We wish to find one such kinetic term.

To determine what valid kinetic term we should introduce for $f_{\mu
  \nu}$, it is first helpful to rewrite the above mass term so that it
more closely resembles a potential for $f_{\mu \nu}$.  Using
$\sqrt{-\det g} = \sqrt{-\det f} \, \sqrt{\det f^{-1}g}$, we have, \be
\label{mass}
\sqrt{-\det g}\, \sum_{n=0}^4 \beta_n \, e_n(\sqrt{g^{-1} f}) =
\sqrt{-\det f} \, \, e_4(\sqrt{f^{-1}g}) \, \sum_{n=0}^4 \beta_n \,
e_n(\sqrt{g^{-1} f}) \, . 
\ee
We would also like to express the $e_n(\sqrt{g^{-1} f})$ in terms of
the inverse matrix $\sqrt{f^{-1}g}$.  Note that the elementary
symmetric polynomials of $\sqrt{f^{-1}g}$ can be easily expressed in
terms of the eigenvalues of $\sqrt{g^{-1}f}$,
\bea
\label{e2}
e_0(\sqrt{f^{-1}g})&=&1, \\[.1cm] 
e_1(\sqrt{f^{-1}g})&=&\frac{1}{\lambda_1}+\frac{1}{\lambda_2}+
\frac{1}{\lambda_3}+\frac{1}{\lambda_4}  \, , \nonumber \\[.1cm] 
e_2(\sqrt{f^{-1}g})&=&\frac{1}{ \lambda_1 \lambda_2} + 
\frac{1}{\lambda_1 \lambda_3} + \frac{1}{\lambda_1 \lambda_4}+
\frac{1}{\lambda_2 \lambda_3} + \frac{1}{\lambda_2 \lambda_4 }+
\frac{1}{\lambda_3 \lambda_4}  \, , \nonumber   \\[.1cm] 
e_3(\sqrt{f^{-1}g})&=&\frac{1}{ \lambda_1 \lambda_2 \lambda_3}+ 
\frac{1}{\lambda_1 \lambda_2 \lambda_4}+\frac{1}{\lambda_1 \lambda_3
\lambda_4}+\frac{1}{\lambda_2 \lambda_3 \lambda_4}\,,\nonumber
\\[.1cm]   
e_4(\sqrt{f^{-1}g})&=&\frac{1}{\lambda_1\lambda_2\lambda_3\lambda_4}
\, . \nonumber  
\eea
Comparing (\ref{e1}) and (\ref{e2}), it is evident that the
$e_n(\sqrt{g^{-1}f})$ can be expressed entirely in terms of the
$e_n(\sqrt{f^{-1}g})$,
\be 
e_k(\sqrt{g^{-1}f})=\frac{e_{4-k}(\sqrt{f^{-1}g})}{e_4(\sqrt{f^{-1}g})}\,.  
\ee 
Then it follows that the mass term (\ref{mass}) can be re-expressed as,
\be
\sqrt{-\det g}\, \sum_{n=0}^4 \beta_n \, e_n(\sqrt{g^{-1} f})
=\sqrt{-\det f}\,\sum_{n=0}^4\beta_n \, e_{4-n}(\sqrt{f^{-1}g})\,. 
\ee
Now it is apparent that the mass term does in fact have the same
ghost-free structure for $f_{\mu \nu}$ as for $g_{\mu \nu}$, only with
different coefficients.  This implies that a valid kinetic term for
$f_{\mu \nu}$ is the standard Einstein-Hilbert kinetic term after all.
Introducing such dynamics for $f_{\mu \nu}$ will give a theory that we
expect to be free of the Boulware-Deser ghost in both the $g_{\mu
  \nu}$ and $f_{\mu \nu}$ sector,
\bea
\label{bimetric}
S&=&M_g^2\int d^4x\sqrt{-\det g}\,R^{^{(g)}}+M_f^2\int d^4x\sqrt{-\det
  f}\,R^{^{(f)}} \nonumber \\ 
&&+2m^2 M_{\rm eff}^2 \int d^4x\sqrt{-\det g}\sum_{n=0}^{4} \beta_n\,
e_n(\sqrt{g^{-1} f}) \, . 
\eea
Here $R^{^{(g)}}$ denotes the scalar curvature for $g_{\mu \nu}$ and $R^{^{(f)}}$ denotes the scalar curvature for $f_{\mu \nu}$. We have introduced
different Planck masses for the two metrics and defined an
``effective'' Planck mass as,
\be
M^2_{\rm eff} =\Big(\frac{1}{M_g^2}+ \frac{1}{M_f^2}\Big)^{-1} \, .
\ee
Given the arguments of \cite{dRG2,dRGT,HR3,HR4}, the mass term
(\ref{mass}) is the unique interaction term that avoids the
Boulware-Deser ghost in the $g$-sector with a non-dynamical $f$.
Here we have shown that it also avoids the ghost in the $f$-sector
with a non-dynamical $g$. 

This structure suggests that the bimetric action is free of the
pathological Boulware-Deser modes.  However, when both $g$ and $f$ are
dynamical, it is not yet obvious how the mass term avoids the two
Boulware-Deser ghosts simultaneously.  In addition, the above arguments are
not sufficient to determine the total number of propagating modes of
the theory at the non-linear level. We examine these issues in the
following sections.

As for the uniqueness of the kinetic term for $f_{\mu \nu}$, it remains to be seen if
there exist other such terms, or if consistent kinetic mixings can be introduced
between $g_{\mu\nu}$ and $f_{\mu\nu}$. However, given the theorem of
\cite{BDGH}, such terms must vanish in the limit $m \rightarrow 0$,
while higher derivative mixings can be neglected compared to the
potential in the long-wavelength limit \cite{DK}. Consequently, we
consider only the action given by (\ref{bimetric}) in what follows.

\section{Spectrum of the linear theory}

Bimetric theories invariably describe one massive and one massless spin-2 field \cite{ISS,BD,SS,DK}.  This is apparent for any theory which reduces to the Fierz-Pauli theory \cite{FP1,FP2} at the linear level.  There, the fluctuation $g_{\mu\nu}-f_{\mu\nu}$ acquires a mass as this is the only combination of metrics that appears in the potential. 

To see this explicitly, let us consider small perturbations in the
minimal massive model introduced in \cite{HR1}. The interaction term
of the minimal model is given by 
\be
\label{actmin}  
2m^2 M_{\rm eff}^2\,\int d^4x\sqrt{-\det g}\left(3-\tr
\sqrt{g^{-1}f}+\det \sqrt{g^{-1}f}\right) \, ,  
\ee
corresponding to the coefficients,
\be
\beta_0=3\, , ~~\beta_1=-1\, , ~~\beta_2=0\, , ~~\beta_3=0\, , ~~\beta_4=1\, .
\ee
Let us expand both $g_{\mu \nu}$ and $f_{\mu \nu}$ around the same
fixed background $\bar{g}_{\mu \nu}$, 
\be
g_{\mu \nu} = \bar{g}_{\mu \nu}+\frac{1}{M_g} h_{\mu \nu} \, ,~~~~
f_{\mu \nu} = \bar{g}_{\mu \nu}+\frac{1}{M_f} l_{\mu \nu} \, .~~~~
\ee
To second order in perturbations, the minimal model gives
\bea
\label{Slin}
S&=&\int d^4x\,(h_{\mu \nu} \hat{{\cal E}}^{\mu\nu\alpha\beta}h_{\alpha \beta}
+l_{\mu \nu} \hat{{\cal E}}^{\mu\nu\alpha\beta}l_{\alpha \beta})\\
&&-\frac{m^2 M_{\rm eff}^2}{4} \int d^4x\,\left[\left(\frac{h^\mu_{~\nu}}{M_g}
-\frac{l^\mu _{~\nu}}{M_f}\right)^2
-\left(\frac{h^\mu_{~\mu}}{M_g}-\frac{l^\mu_{~\mu}}{M_f}\right)^2\right] \, . \nonumber
\eea
Here $\hat{{\cal E}}^{\mu\nu\alpha\beta}$ denotes the usual Einstein-Hilbert kinetic operator and indices are raised and lowered with respect to the background metric.

To diagonalize, we perform the change of variables
\bea
\frac{1}{M_{\rm eff}} \,u_{\mu \nu} = \frac{1}{M_f}\,h_{\mu \nu} + \frac{1}{M_g}\,l_{\mu \nu} \, , \\
\frac{1}{M_{\rm eff}} \,v_{\mu \nu} =\frac{1}{M_g}\,h_{\mu \nu} - \frac{1}{M_f}\,l_{\mu \nu} \, .
\eea
The action (\ref{Slin}) becomes
\be
S=\int d^4x\,(u_{\mu \nu} \hat{{\cal E}}^{\mu\nu\alpha\beta}u_{\alpha \beta}
+v_{\mu \nu} \hat{{\cal E}}^{\mu\nu\alpha\beta}v_{\alpha \beta})
-\frac{m^2}{4} \int d^4x\,\left(v^{\mu \nu}v_{\mu \nu} - v^{\mu}_{~\mu}v^{\nu}_{~\nu}\right)\,.
\ee
Thus at the linearized level, we see that the minimal action describes one massless spin-2 particle $u_{\mu \nu}$ and one massive spin-2 particle $v_{\mu \nu}$ with mass $m$.

Note that,
\bea
&{\rm for}~ M_f \gg M_g:&~u_{\mu \nu} \rightarrow l_{\mu\nu}~~{\rm and}~~v_{\mu \nu} \rightarrow h_{\mu\nu}\, , \nonumber \\
&{\rm for}~ M_g \gg M_f:&~u_{\mu \nu} \rightarrow h_{\mu\nu}~~{\rm and}~~v_{\mu \nu} \rightarrow -l_{\mu\nu} \,.\nonumber
\eea
When one Planck mass greatly exceeds the other, it is always the massless particle that has the larger Planck mass.  Thus it is possible to send one Planck mass to infinity so that the massless particle decouples from the theory.  One is left with a formally covariant theory with a single massive spin-2 field propagating in a fixed background.

\section{Constraint analysis: minimal theory}
Beyond the linear level, one can determine the number of degrees of
freedom of the bimetric theory by doing an analysis of the Hamiltonian
and momentum constraints in the ADM formulation \cite{ADM}.  Such an analysis will determine whether or not the
Boulware-Deser ghosts avoided at the linear level reappear in the
non-linear theory.  Let $N$ and $N_i$ denote the lapse and shift functions of the metric $g_{\mu\nu}$ while $L$ and $L_i$ denote the lapse and shift functions of the metric $f_{\mu\nu}$.  Accordingly we define, 
\bea 
N =(-g^{00})^{-1/2}\,, \qquad N_i = g_{0i}\,, \qquad \gamma_{ij}=g_{ij}\,,  \\
L =(-f^{00})^{-1/2}\,, \qquad L_i = f_{0i}\,, \qquad \tf_{ij}=f_{ij}\,.  
\eea 
The canonically conjugate momenta for $\gamma_{ij}$ and $\tf_{ij}$ are
given by $\pi^{ij}$ and $p^{ij}$ respectively.  In the bimetric theory
(\ref{bimetric}) the lapse and shift variables appear without time
derivatives and are thus non-dynamical.  The remaining six $g_{ij}$,
six $f_{ij}$ and their conjugate momenta constitute $24$ phase-space degrees of
freedom, i.e., $12$ potentially propagating modes, a propagating mode
referring to a pair of conjugate variables.

Before before proceeding with the constraint analysis of the minimal model we note that, based on the arguments of the previous sections, we expect the counting of degrees of freedom to be appropriate for a massless spin-2 field (two modes) plus a massive spin-2 field (five modes).  Thus, we need ten total constraints and gauge invariances to leave us with $14$ degrees of freedom, corresponding to seven total propagating modes.  The general covariance should be responsible for removing eight degrees of freedom.  Four degrees of freedom will be removed by gauge fixing while the other four will be removed via the associated Hamiltonian and momentum constraints.  Indeed, we will show below that for the minimal theory, one set of lapse and shift variables appear linearly, thus their equations of motion act as constraints on the remaining variables.  In addition, we expect there to be a single constraint and associated secondary constraint, as was found in \cite{HR4} and \cite{HR6}.  These remove an additional propagating mode.  Below we will also demonstrate the existence of this constraint in the minimal theory.  Thus, out of $24$ initial degrees of freedom we are left with $14$, consistent with one massive spin-2 field and one massless spin-2 field.

Let us again be more explicit by studying the constraints of the minimal
model (\ref{actmin}).  This action is highly non-linear in the lapse
and shift variables.  However, following \cite{HR4} we introduce three
new variables $n^i$ which are functions of the lapse and shift
variables and spatial metrics,
\be
N^i - L^i  = (L\, \delta^i_{\,k}+ N\, D^i_{\,k}) n^k\,.
\label{NnD}
\ee
The matrix $D =D^i_{\,k}$ is given by,
\be
D=\sqrt{\gamma^{-1}\,\tf \,Q}\,\, Q^{-1}\,,
\label{SolnD}
\ee
where
\be
Q^l_{~j} = \left[1 -n^k\,(\tf_{km})\,n^m\right]\,\delta^l_{~j}
+n^l\,n^m\, (\tf_{mj})\,.
\label{Minv}
\ee
Significantly, $D$ is independent of the lapses $N$ and $L$, and the shifts $N^i$ and $L^i$.  Let us also define 
\be
x \equiv 1 -n^k\,(\tf_{km})\,n^m\,.
\ee

We can use the above variables to replace the shift $N^i$ in favor of the $n^i$.  The minimal action then becomes linear in the lapses $N$ and $L$, and the shift $L^i$, 
\bea
{\cal L}&=&
M_g^2 \pi^{ij}\partial_t \gamma_{ij} +M_f^2 p^{ij}\partial_t \tf_{ij}
+L^i(M_g^2\,R_i^{^{(g)}}+M_f^2\,R_i^{^{(f)}}) \nonumber \\
&&+L\left[M_f^2\,R^{0^{(f)}}\!\!+M_g^2\,n^iR_i^{^{(g)}}-2m^2M^2_{\rm eff}(\sqrt{\det\gamma}\sqrt{x}-\sqrt{\det\tf})\right]\nonumber \\
&&+ N \left[M_g^2\, R^{0^{(g)}}\!\!+M_g^2\, D^i_{\,k}n^k R_i^{^{(g)}} -2m^2M^2_{\rm eff}(\sqrt{\det\gamma}\sqrt{x}D^k_{~k}-3\sqrt{\det\gamma})\right] \, .
\label{Lmin}
\eea
What's more, the equations of motion for the $n^i$ can be shown to be
independent of $N$ \cite{HR4}.  Thus they can be used to fix $n^i$ in
terms of the remaining degrees of freedom,  
\be
n^i=\frac{-M_g^2\,(\tf^{-1})^{ij}\,R_j^{^{(g)}}}{\sqrt{4m^4M^4_{\rm eff}\, 
\det\gamma+M_g^4R_k^{^{(g)}} (\tf^{-1})^{kl}R_l^{^{(g)}}}} \, .
\label{nSoln}
\ee
The $N$ equation of motion is 
\be
M_g^2 R^{0^{(g)}}\!\!+M_g^2 R_i^{^{(g)}} D^i_{\,j}\,n^j
-2m^2M^2_{\rm eff} \sqrt{\det\gamma}\left[
\sqrt{x}\, D^k_{~k} -3 \right]=0 \, .
\label{Neom}
\ee
On substituting for $n^i$ we can use this equation to constrain the
$\pi^{ij}$ and $\gamma_{ij}$. Along with the associated secondary
constraint found in \cite{HR6}, this eliminates one propagating mode. 

A crucial point is that the ghost mode of $\gamma_{ij}$ that is
eliminated through the Hamiltonian constraint (\ref{Neom}) remains
independent of the lapse $L$. Only then does the term $\pi^{ij}\partial_t
\gamma_{ij}$ in the Lagrangian (\ref{Lmin}), when evaluated on the
constraint surface, remain independent of $L$. To see this, note that $n^i$, $\sqrt{x}$ and $D$ are independent of the lapse $L$.  Thus, the
Hamiltonian constraint (\ref{Neom}) and the associated secondary
constraint are independent of $L$ as well.  Using the constraint
(\ref{Neom}), the would-be ghost mode of the $g$-sector can be
expressed in terms of the remaining independent variables.  It follows
that the expression for the would-be ghost mode does not involve $L$
either.

Now we use equations (\ref{nSoln}) and (\ref{Neom}) to eliminate
the lapse $N$ and the new shift-like variables $n^i$ of the $g$-metric
from the Lagrangian (\ref{Lmin}).  The resulting expression remains linear in both the lapse $L$ and shift $L^i$ of the $f$-metric,
\bea
{\cal L}&=& M_g^2\,\pi^{ij}\partial_t \gamma_{ij} + M_f^2\,p^{ij}\partial_t \tf_{ij} 
+L^i\left(M_g^2\,R_i^{^{(g)}}+M_f^2\,R_i^{^{(f)}}\right)  \\[.2cm]
&&+L\,\Bigg\{M_f^2\,R^{0^{(f)}}\!\!- 2m^2M^2_{\rm eff} \Bigg[\sqrt{\det \gamma+ 
\frac{M_g^4}{4 m^4M^4_{\rm eff}}\,R_k^{^{(g)}}(\tf^{-1})^{kl} R_l^{^{(g)}}}-\sqrt{\det\tf}\Bigg] \Bigg\}\,.   \nonumber
\label{Lmin2}
\eea
Thus $L^i$ and $L$ are Lagrange multipliers.  Their equations of
motion act as four constraints on the six canonical pairs
$(p^{ij},\tf_{ij})$ and the remaining five independent pairs of $(\pi^{ij},\gamma_{ij})$,  
\be
\label{P-f}
M_g^2\,R_i^{^{(g)}} \!\!+ M_f^2\,R_i^{^{(f)}}\!\!= 0\, ,
\ee
\be
\label{H-f}
\frac{M_f^2}{2m^2M^2_{\rm eff}}\,R^{0^{(f)}}\!\!
 -\sqrt{\det \gamma+\frac{M_g^4}{4 m^4M^4_{\rm eff}}\, R_k^{^{(g)}}(\tf^{-1})^{kl}R_l^{^{(g)}}}+\sqrt{\det\tf}=0\,.
\ee
Note that using (\ref{P-f}), we can rewrite (\ref{H-f}) in terms of the $f_{\mu\nu}$ variables $R^{^{(f)}}$ as
\be
\label{H-f2}
\frac{M_f^2}{2m^2M^2_{\rm eff}}\,R^{0^{(f)}} \!\!
-\sqrt{\det \gamma+\frac{M_f^4}{4 m^4M^4_{\rm eff}}\, R_k^{^{(f)}}(\tf^{-1})^{kl} R_l^{^{(f)}}}+\sqrt{\det\tf} =0\,.
\ee

The situation now resembles that of general relativity. These four
constraints along with four coordinate conditions can be used to
eliminate another four pairs of canonical variables, including the ghost
in the $f$-sector. Thus we are indeed left with seven propagating
modes, consistent with one massive and one massless spin-2 field. 
Roughly speaking, (\ref{H-f}) can be used to eliminate the
Boulware-Deser mode of the $f$-sector while the constraints (\ref{P-f}) eliminate one
combination of the helicity-1 and helicity-0 parts of $g_{\mu\nu}$ and $f_{\mu\nu}$. The
surviving combination, together with the corresponding combination of 
helicity-2 modes of $g_{\mu\nu}$ and $f_{\mu\nu}$, will make up the massive state.
After fixing the coordinate conditions, the $L$ and $L^i$ are determined by four of the $\tf_{ij}$ and $p^{ij}$
equations of motion.  Another four of these equations should be trivially satisfied as a
consequence of Bianchi identities.

Note that we have introduced our new shift-like variables (\ref{NnD}) in such a way that an asymmetry arises between the $g$-sector and the $f$-sector in the ADM analysis.  Namely, there appears to be only one constraint arising from the $g$-sector of the theory while there are four constraints in the $f$-sector.  Let us emphasize that this apparent asymmetry is purely a result of the choice of variables (\ref{NnD}) and that no fundamental asymmetry exists between these two sectors, as is evident from the constraint equations (\ref{Neom}), (\ref{P-f}) and (\ref{H-f2}).

\section{Constraint analysis: full theory}
Let us now extend our analysis to the full bimetric theory (\ref{bimetric}).  We write this action in terms of the ADM variables and again replace the shift $N^i$ with $n^i$ as defined in (\ref{NnD}).  The full Lagrangian is then linear in the lapses $N$ and $L$, and the shift $L^i$,
\be
{\cal L} = M_g^2 \, \pi^{ij}\partial_t \gamma_{ij}+M_f^2 \, p^{ij}\partial_t \tf_{ij}
 - {\cal H}_0+N {\cal C}\, ,
\label{Lbi}
\ee
where ${\cal H}_0$ and ${\cal C}$ stand for 
\begin{align}
{\cal H}_0 &= -L^i\left(M_g^2 R_i^{^{(g)}}\!\!+M_f^2 R_i^{^{(f)}}\right)
-L\left(M_f^2 R^{0^{(f)}}\!\!+ M_g^2 n^iR_i^{^{(g)}} \!\!+2m^2 M^2_{\rm eff} \sqrt{\det \gamma} \, U \right) \, ,
\\[.1cm]
{\cal C} &= M_g^2 R^{0^{(g)}}\!\!+ M_g^2  R_i^{^{(g)}}D^i_{\,j}n^j +2m^2 M^2_{\rm eff} \sqrt{\det \gamma} \, V\,. \label{Cbi}
\end{align}
$U$ and $V$ in the above expressions are defined as 
\bea
U &\equiv &\beta_1 \sqrt{x} 
+\beta_2(\sqrt{x}^{\,2}\,D^i_{~i}+n^i \,\tf_{ij}D^j_{~k}n^k)  \\
&&+\beta_3\left[\sqrt{x}\, (D^l_{~l}\,n^i \, \tf_{ij}D^j_{~k}n^k-D^i_{~k}\,n^k \, \tf_{ij}D^j_{~l}n^l)
+\tfrac{1}{2}\sqrt{x}^{\,3}(D^i_{~i}D^j_{~j}-D^i_{~j}D^j_{~i}) \right] \nonumber \\
&&+\beta_4 \sqrt{\det \tf}/\sqrt{\det \gamma} \, , \nonumber \\
V &\equiv &\beta_0+\beta_1 \sqrt{x} \,D^i_{~i}
+\tfrac{1}{2}\beta_2\sqrt{x}^{\,2}\,(D^i_{~i}D^j_{~j}-D^i_{~j}D^j_{~i})  \\
&&+\tfrac{1}{6}\beta_3\sqrt{x}^{\,3}\,(D^i_{~i}D^j_{~j}D^k_{~k}-
3D^i_{~i}D^j_{~k} D^k_{~j}+2D^i_{~j} D^j_{~k} D^k_{~i} ) \, . \nonumber
\eea
Varying the Lagrangian with respect to $N$ gives the equation of motion,
\be
\label{C}
{\cal C} = 0 \, ,
\ee
while the $n^i$ equation of motion is given by \cite{HR4}
\bea
\label{Ci}
{\cal C}_i &\equiv& M_g^2\,R_i^{^{(g)}}-2m^2M^2_{\rm eff}\sqrt{\gamma}\,\frac{n^lf_{lj}}{\sqrt{x}}\bigg[\beta_1\,
\delta^j_{i} +\beta_2\,\sqrt{x}\, (\delta^j_{i} D^m_{~m}-D^j_{~i})\nonumber \\
&&+\beta_3 \, \sqrt{x}^{\,2} \, \left(\tfrac{1}{2}\delta^j_{~i} 
(D^m_{~m} D^n_{~n}\, -\,D^m_{~n} D^n_{~m}) +D^j_{~m} D^m_{~i}\, -\,
D^j_{~i} D^m_{~m}\right) \bigg] = 0 \, .
\eea
This equation is independent of the lapses $N$ and $L$, and the shift $L^i$ and can in principle be used to determine the $n^i$ in terms of the dynamical variables $\gamma_{ij}$ and $\pi^{ij}$.  Thus the $N$ equation (\ref{C}) becomes a constraint on the dynamical variables, reducing the 12 degrees of freedom to 11.  This is the Hamiltonian constraint.

In addition, because $n^i$, $\sqrt{x}$ and $D$ are independent of the lapse $L$ and shift $L^i$, the Lagrangian (\ref{Lbi}) remains linear in $L$ and $L^i$ after the $n^i$ have been eliminated in favor of $\gamma_{ij}$ and $\pi^{ij}$.  Thus the lapse and shift remain Lagrange multipliers, enforcing the correct number of constraints, along with the general coordinate invariances, and primary and secondary constraints, for one massive and one massless spin-2 field. 

\section{Coupling to matter}
The coupling of matter to the metrics $g_{\mu\nu}$ and $f_{\mu\nu}$
must not violate the constraints found above.  Hence the coupling
should be linear in the lapse and shift functions of both $g_{\mu \nu}$ and $f_{\mu\nu}$. 
The minimal coupling of General Relativity to matter,
$\sqrt{-\det g} \,{\cal L}_M$, automatically satisfies this
requirement.  If this were not the case, the constraints of even massless GR would be violated.  Thus the simplest allowed matter couplings are of the
form,    
\be
\sqrt{-\det g} \, {\cal L}_1(g_{\mu\nu}, \phi^A) \, ,~\sqrt{-\det f}
\, {\cal L}_2(f_{\mu\nu}, \phi^B) \, . 
\ee
Here the $\phi^{A,B}$ represent any matter field.  

If both $g_{\mu\nu}$ and $f_{\mu\nu}$ are allowed to couple to the same matter fields, the particle geodesic equations will be modified, generically violating the equivalence principle.  Thus in many cases such couplings could be easily ruled out experimentally.  This problem could be avoided if matter couples to an effective metric $\hat g_{\mu\nu}$ constructed out of $g_{\mu\nu}$ and $f_{\mu\nu}$, as 
\be
\sqrt{-\det\hat g} \,\, {\cal L}_3(\hat g_{\mu\nu}, \phi^C)\, .
\ee
However, such a coupling will generically violate the constraints. Thus the allowed couplings to matter in the bimetric theory are severely restricted.  The observational implications of these couplings and possible classical solutions will be discussed in future work.

\vskip.5cm

\acknowledgments

We would like to thank especially G. Gabadadze and K. Hinterbichler for helpful discussions. The majority of this
work was completed while R.A.R. was supported by the Swedish Research Council (VR) through the Oskar Klein Centre.  R.A.R. is currently supported by NASA under contract NNX10AH14G.

\end{document}